\documentclass{jfm}

\usepackage{graphicx}
\usepackage{newtxtext}
\usepackage{newtxmath}
\usepackage{natbib}
\usepackage{hyperref}
\hypersetup{
    colorlinks = true,
    urlcolor   = blue,
    citecolor  = black,
}

\newcommand{\RomanNumeralCaps}[1]
\linenumbers

\newcommand{\Roneb}[1]{#1}
\newcommand{\Rtwob}[1]{#1}
\newcommand{\Rthreeb}[1]{#1}
\newcommand{\Rusb}[1]{#1}

\newcommand{\Rone}[1]{#1}
\newcommand{\Rtwo}[1]{#1}
\newcommand{\Rthree}[1]{#1}
\newcommand{\Rus}[1]{#1}

\newcommand{\lderiv}[2][]{\frac{\text{D}#2}{\text{D}t}}
\newcommand{\deriv}[3][]{\frac{\text{d}#2}{\text{d}#3}}
\newcommand{\dfm}[2][]{\overline{#2}}

\newcommand{\pdel}[2]{{#1}_{\!,{#2}}}
\newcommand{\di}[0]{\,\text{d}}

\title{Near-inertial echoes of ageostrophic instability in submesoscale filaments}

\author{Erin Atkinson\aff{1}
  \corresp{\email{erin.atkinson@mail.utoronto.ca}},
  James C.\ McWilliams\aff{2} and Nicolas Grisouard\aff{1}}

\affiliation{\aff{1}Department of Physics, University of Toronto, 60 St. George Street, Toronto, Ontario, Canada M5S 1A7\aff{2}Department of Atmospheric and Oceanic Sciences, University of California, Los Angeles, Los Angeles, CA, USA}

\begin{document}
\maketitle

\begin{abstract}
Ocean submesoscales, flows with characteristic size around $10~\text{m}-10~\text{km}$, are transitional between the larger, rotationally-constrained mesoscale and three-dimensional turbulence. In this paper we present simulations of a submesoscale ocean filament. In our case, the filament is strongly sheared in both vertical and cross-filament directions and is unstable. Instability indeed dominates the early behaviour with a fast extraction of kinetic energy from the vertically sheared thermal wind. \Rthree{However, the} instability \Rthree{that} emerges does not exhibit characteristics that match \Rtwob{the perhaps expected symmetric or Kelvin-Helmholtz instabilities}\Rthree{, and appears to be non-normal in nature.} \Roneb{The prominence of the transient response depends on the initial noise and, for large initial noise amplitudes, saturates before SI normal modes are able to develop.} The action of the instability is sufficiently rapid --- \Rone{with energy extraction from the mean flow emerging and peaking within the first inertial period ($\sim 18~\text{hr}$)} --- that the filament does not respond in a geostrophically balanced sense. Instead, \Rtwob{at all initial noise levels}, it later exhibits vertically sheared near-inertial oscillations with higher amplitude as the initial minimum Richardson number decreases. Horizontal gradients strengthen only briefly as the fronts restratify. These unstable filaments can be generated by strong mixing events at pre-existing stable structures; \Rone{we also caution against inadvertently triggering this response in idealised studies that start in a very unstable state.}
\end{abstract}



\section{Introduction}


\Rthree{At length scales larger than $\sim 100 \text{ km}$, flows in the ocean are close to geostrophic balance --- a state of approximate balance between the Coriolis acceleration and pressure gradient --- and evolve slowly, over many days. At smaller length scales, beyond those that can currently be resolved in global ocean models, lies the submesoscale ($\sim 0.1- 10\text{ km}$).} \Rone{Understanding of the impact of submesoscale flows on the ocean is emerging, with observational and numerical evidence that processes at this scale, which is poorly represented in parametrisations, are crucial components of the ocean's physical and biogeochemical processes \citep{Mahadevan2016, Calil2021, gula2022submesoscale, Aravind2023, Pham2024, Zhu2024}.}

\Rthree{Images of ocean properties at submesoscale resolution are often peppered with highly anisotropic features. Fronts --- confined lateral gradients in density --- and filaments --- opposing pairs of fronts --- remain balanced by the Coriolis acceleration of an along-front jet. In the open ocean, these features are created primarily at the edges of mesoscale vortices such as those created by separating western boundary currents, such as the Gulf Stream or Kuroshio current \citep{2016_PRSA_McWilliams}.} As \Roneb{fronts and filaments} are drawn out by the strain flow in these regions, horizontal gradients in velocities and tracers become larger, eventually leading to the onset of a new dynamical regime, where geostrophic and hydrostatic balances no longer describe the flow well \Rtwo{\citep{Shakespeare2016, 2019_JPO_BarkanMSMD, 2023_ARFM_TaylorThompson}}.

\Rtwo{In addition to the along-front geostrophic jet, an ageostrophic secondary circulation may develop at a front due to a background strain, turbulent mixing, instability or otherwise \citep{2018_JFM_SullivanMcWilliams, 2018_JFM_CroweTaylor, Mcwilliams2021}.} The resulting vertical velocities provide a route for transport of tracers, momentum and energy between the surface layer and deep water \citep{Mahadevan2016, Klein2019, 2019_OM_VermaPS, Freilich2021}. \Rtwob{In particular}, the submesoscale is a link between the larger scale, horizontally non-divergent flows and smaller scale turbulence that facilitates the observed forward cascade of energy in the ocean \citep{Srinivasan2023}.

In the absence of turbulence, \Rtwob{finite time singularities form in the horizontal gradients} when forced by a strain flow \citep{Hoskins1972}. For a realistic front, some effective horizontal diffusion due to turbulence will arrest the frontogenesis before this singularity is reached, though this process is not well-understood \Rtwo{\citep{samelson2016frontogenesis, 2019_OM_VermaPS}}. The turbulence, namely the vertical momentum mixing induced by surface cooling or wind stress, acts on the vertically sheared thermal wind velocity. If only these processes act in the submesoscale mixed layer, the expected momentum balance is the \textit{turbulent thermal wind} balance between the Coriolis acceleration of an along-front jet, the horizontal pressure gradient and vertical momentum mixing \Rtwo{\citep{2014_JPO_GulaMM, 2018_JFM_CroweTaylor, Bodner2023}}.
 
Of particular interest in this study is symmetric instability (SI). A symmetrically unstable flow is unstable to perturbations that exchange fluid parcels along isopycnals, with no variation along the direction of mean flow \citep[hence symmetric;][]{1966_JAS_Stone}. The instability manifests as highly anisotropic rolls, along the direction of the frontal jet, with major axis aligned with isopycnals. \Rone{The most unstable across-front wavenumber tends to infinity in the inviscid limit, so the wavenumber of the observed mode is presumed to be restricted by diffusion \citep{Harris2022}.} SI extracts kinetic energy from the thermal wind velocity shear \Rtwo{\citep{1998_JPO_HaineMarshall, Thomas2013, 2023_ARFM_TaylorThompson}}.

For a Boussinesq fluid with equal momentum and buoyancy diffusivities, the stability of a flow to symmetric perturbations can be diagnosed by the sign of Ertel potential vorticity $q$. Fluid unstable to SI has $fq = f(f\hat {\boldsymbol{z}} + \nabla \times \boldsymbol{u}) \cdot \nabla b < 0,$ where $\boldsymbol u = u \hat {\boldsymbol{x}} + v \hat {\boldsymbol{y}} + w \hat {\boldsymbol{z}}$ and $b$ are velocity and buoyancy fields respectively and $f$ is the Coriolis frequency \citep{Emanuel1979}. SI competes with gravitational, inertial, baroclinic and Kelvin-Helmholtz instabilities, and dominates in regions where the Richardson number \Rtwo{($Ri = N^2 / S^2$ where $N$ and $S$ are buoyancy frequency and vertical shear of horizontal velocity respectively)} of the flow lies in the range $0.25 < Ri < 0.95$ \citep{1966_JAS_Stone, Thomas2013}. The small size and strong baroclinicity of submesoscale fronts often \Rus{produces favourable conditions for SI} \citep{Thomas2013, 2014_JPO_GulaMM}.

\Rone{The behaviour of fluid susceptible to SI (or close to unstable) is suspected to have important consequences for ocean flows at the submesoscale and, for example, has been studied in the contexts of frontogenesis and frontal arrest \citep{Thomas2012, 2019_OM_VermaPS}, mixing and vertical transport \citep{Brannigan2016, Bosse2021}, and near-inertial wave generation \citep{Wienkers2021}.} \Rthree{However, the initial and long-time behaviour of a symmetrically unstable flow differ in form, growth rate and energy source. \Rthreeb{The form of the linear perturbations that optimally extract energy from an Eady background state differ from the most unstable normal mode during initial evolution and their growth rate is enhanced during an initial, transient phase of instability. This behaviour persists in fully non-linear, three-dimensional simulations} \citep{2007_JAS_HeifetzFarrell, zemskova2020transient, kimura2024initial}. Initial motions in an unstable flow seeded with noise do not resemble the symmetric normal modes, and exhibit a faster growth rate across parameter space, a behaviour that is general across hydrodynamic stability \citep{1993_Science_TrefethenTRD, zemskova2020transient}.
This non-normal, transient growth will play a particularly important role in this study.}

In this paper, we present a set of \Rus{strongly baroclinic} ($Ri \leq 0.2$) simulations \Rthree{resembling the configuration of \citet{2018_JFM_SullivanMcWilliams}, but with coarser spatial resolution and without the surface cooling that drives vertical mixing. \Roneb{On the other hand, we run more simulations, allowing for a better parameter space exploration, and we do so for longer periods of time, namely, several inertial periods each time.}}

\Roneb{As in the aforementioned study, the reference state is unstable; and we pay particular attention to the consequences of instability when initial noise is applied and how this may obfuscate conclusions that assume a balanced turbulent thermal wind state.}
\Roneb{Specifically, we share the goal of \citet{2018_JFM_SullivanMcWilliams} to have realistic levels of noise, but while they attempt to generate realistic turbulent noise by continuously forcing the ocean surface, we simply generate random noise that aims to mirror the turbulent kinetic energy profile of mixed layer convection driven by surface cooling. This also results in a larger amplitude than many idealised studies, which seed the fluid with very small levels of noise \citep{zemskova2020transient, Wienkers2021_2}.}

We show that instabilities act quickly to remove kinetic energy and excite near-inertial oscillations within the fronts that briefly strengthen horizontal buoyancy gradients at the surface. \Rthree{We argue that we observe a transient response of the filament \Roneb{as it evolves towards normal-mode} symmetric instability.} Finally, we suggest that \Roneb{the initialisation of both turbulence and an unstable reference state must be done with care and} \Rthree{initial conditions of} numerical studies of similar strongly baroclinic features must be designed with initial instability and possible subsequent oscillations in mind, or risk drawing conclusions that may not be generalisable to balanced flows in the real ocean. 
\section{\Rus{Numerical simulations}}
\label{sec:simulations}
The simulations presented in this paper are of a set of buoyancy filaments in a rotating, Boussinesq fluid, namely,
\refstepcounter{equation}
$$
\lderiv{\boldsymbol u} + f \boldsymbol{\hat z} \times \boldsymbol u = -\nabla \varphi + b\boldsymbol{\hat z}  + \nabla \cdot (\nu \nabla \boldsymbol{u})+ \boldsymbol F_{\boldsymbol u},\eqno{(\theequation{a})}\label{eq:boussinesq}$$
$$
\lderiv{b} = \nabla \cdot (\nu \nabla b) + F_b, \quad \text{and} \quad 
        \nabla \boldsymbol{\cdot} \boldsymbol u = 0,  \eqno{(\theequation{\mathit{b},\mathit{c}})}
$$
with \Rone{$f = 10^{-4}~\text{s}^{-1}$}, geopotential $\varphi$ and \Rtwo{Lagrangian derivative ${\text{D}\phi}/{\text{D}t} = \partial \phi / \partial t + \boldsymbol{u} \cdot \nabla \phi$.} The terms $\boldsymbol{F_u}$ and $F_b$ are velocity and buoyancy forcing terms. We elaborate on these quantities below. The equation set \eqref{eq:boussinesq} is integrated using Oceananigans, a Julia package for simulation of incompressible fluid flows in an oceanic context \citep{Ramadhan2020}.
\begin{figure}
    \centering
    \includegraphics[width=\textwidth]{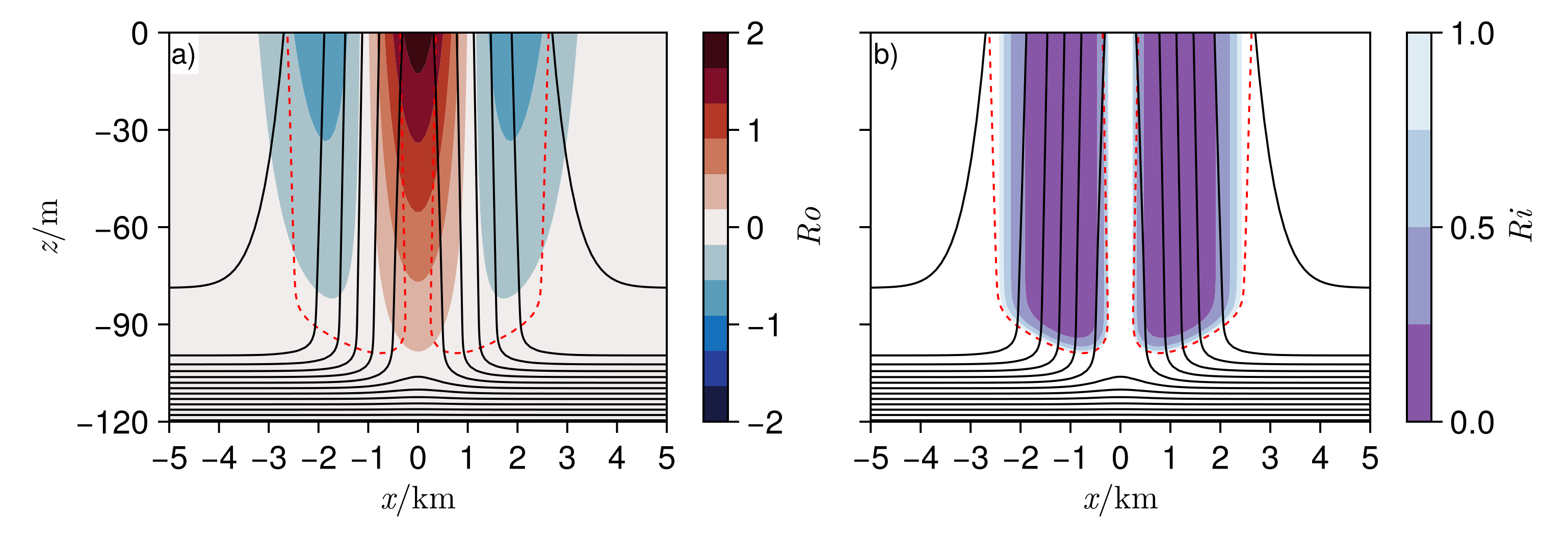}
    \caption{\Rtwo{a) The local Rossby number for the reference state with $Ri_{\min{}} = 0.1$. Black contours are isopycnals. Fluid inside the red, dashed contour has $fq < 0$.} b) The local Richardson number, as in a).}
    \label{fig:schematic}
\end{figure}
The grid size is $(N_x, N_y, N_z) = (1024, 1024, 128)$ and the simulation domain is $(x, y)\in [-5L, 5L)^2$, \Rone{where $L=1000\text{ m}$} is a measure of the initial surface separation between the centres of the two counter-flowing jets, and $z\in [-2.5H, 0]$, \Rone{where $H=100\text{ m}$} is the initial depth of the mixed layer. The grid spacing is uniform in the horizontal directions. The vertical grid spacing smoothly transitions between two uniform values at the pycnocline, such that approximately $3/4$ of the cells are in the surface mixed layer ($z>-H$), while the remainder are in the deep water.

The domain is periodic in both horizontal directions. In the vertical, boundary conditions are no-flux ---  the filament is not forced by wind stress or surface cooling, namely, \Rtwo{$u_{,z} = v_{,z} = w = b_{,z} = 0$ at $z=0$ and $u_{,z} = v_{,z} = w = 0$ and $b_{,z} = N_0^2$ at $z=-2.5H$,} where $\phi_{,i}$ denotes differentiation of $\phi$ with respect to coordinate $i$ and \Rtwo{$N_0$ is the deep water buoyancy frequency.}

To prevent internal wave reflections from the bottom boundary, a sponge layer forcing $\boldsymbol F_{\boldsymbol u}, F_b$ is included for all fields. The fields are relaxed towards the reference state ($\boldsymbol u_0, b_0$) at a rate that decays quadratically away from the bottom boundary, namely,
\refstepcounter{equation}
$$
  \boldsymbol F_{\boldsymbol u} = - \sigma(z) (\boldsymbol u - \boldsymbol u_0) \quad \text{and} \quad
  F_b = -\sigma(z) (b - b_0)
  \eqno{(\theequation{\mathit{a},\mathit{b}})}\label{eq:forcing}
$$
\begin{equation}
    \text{with}\quad \sigma(z) = \frac12\frac{N_0}{2\pi}\times \left \{\begin{array}{ll}
        (2z/H + 4)^2, & z \leq -2H, \\
        0, & z > -2H.
    \end{array}\right .
\end{equation}

The isotropic eddy viscosity $\nu$ is parametrised as a Smagorinksy-Lilly turbulent closure with a Smagorinsky constant $C=0.16$ \citep{smagorinsky1963general, lilly1967representation}, and we set $\kappa = \nu$. \Rtwo{The adaptive time-stepping routine aims to achieve a maximum Courant-Friedrichs-Levy number of 0.5}. 

The reference state of the simulation is a buoyancy profile $b_0(x, z)$, and a jet $\boldsymbol u_0 = v_0(x, z) \boldsymbol{\hat y}$ in thermal wind balance, namely,
$$
    v_0 = \frac{1}{f}\int_{-\infty}^z \frac{\partial b_0}{\partial x} \; \text{d}z.
$$
The reference buoyancy contains a weakly-stratified surface layer above a strongly-stratified fluid. The height of the transition between these two stratifications varies, providing the horizontal buoyancy gradient that makes up the filament. \Rone{The reference filament state is shown in figure \ref{fig:schematic}}, \Rtwo{and the exact definition of the functional forms of $v_0$ and $b_0$ are given in appendix \ref{app:filament}.}

The reference state is invariant in the down-front direction $y$ and the initial evolution introduces little variation along this direction for the large-scale velocity and buoyancy fields (i.e. the filament does not meander \Rtwob{as the short down-front domain constrains the growth rate of baroclinic instabilities}). As such, we define the mean state to simply be the down-front average of the full 3D state, with averaging operator
\begin{equation}
    \dfm{\phi}(x, z, t) := \frac{1}{10L}\int_{-5L}^{5L} \phi(x, y, z, t) \;\text{d}y,
    \label{eq:dfm}
\end{equation}
and down-front \Rtwo{fluctuations $\phi' = \phi - \dfm{\phi}$}.
The primary non-dimensional numbers that determine the dynamical regime of an inviscid submesoscale mixed layer flow are the Rossby and Richardson numbers \citep{2016_PRSA_McWilliams, Bodner2023}. For the base state, the local values of each are, respectively,
\begin{gather}
    Ro = \frac{1}{f}\frac{\partial v_0}{\partial x},\quad\text{and}\quad Ri = \frac{\partial b_0 / \partial z}{(\partial v_0 / \partial z)^2}.
\end{gather}
We parametrise our initial conditions by extremal values of the local Rossby and Richardson numbers as follows.

This study is concerned with the submesoscale ocean, where advection and the Coriolis acceleration are comparable in magnitude: $Ro \sim 1$. To limit the number of possible instabilities, the \Rone{reference state} has minimum vertical vorticity above $-f$, which excludes inertial instability. We use $Ro_\text{min} = -0.8$ here. Given the prescribed filament shape, this results in a maximum $Ro_\text{max} = 1.9$. \Rone{For typical horizontal velocities in submesoscale fronts this corresponds to jet widths of order $1-10~\text{km}$ at mid-latitudes \citep{Siegelman2020, Calil2021}.}

The submesoscale mixed layer often has $Ri \lesssim 1$ \Rtwo{\citep{Zhu2024}}. For the filament shape used here, the Richardson number is approximately constant with depth in the mixed layer, and attains a minimum value in the \Rus{horizontal} centre of each front. We present the results of a set of low-Ri filaments, with a range of minimum Richardson numbers: $Ri_\text{min} \in \{0.0, 0.1, 0.2\}$. The remainder of the non-dimensional parameters, as well as the relationship between $Ro_\text{min}$, $Ri_\text{min}$ and dimensional parameters, is given in appendix \ref{app:filament}. Figures \ref{fig:schematic}a and \ref{fig:schematic}b show the local \Rtwo{$Ro$} and $Ri$ of the $Ri_{\min{}} = 0.1$ reference state respectively, and the region with $fq<0$ at the outer edges of the filament.

\begin{figure}
    \centering
    \includegraphics[width=\textwidth]{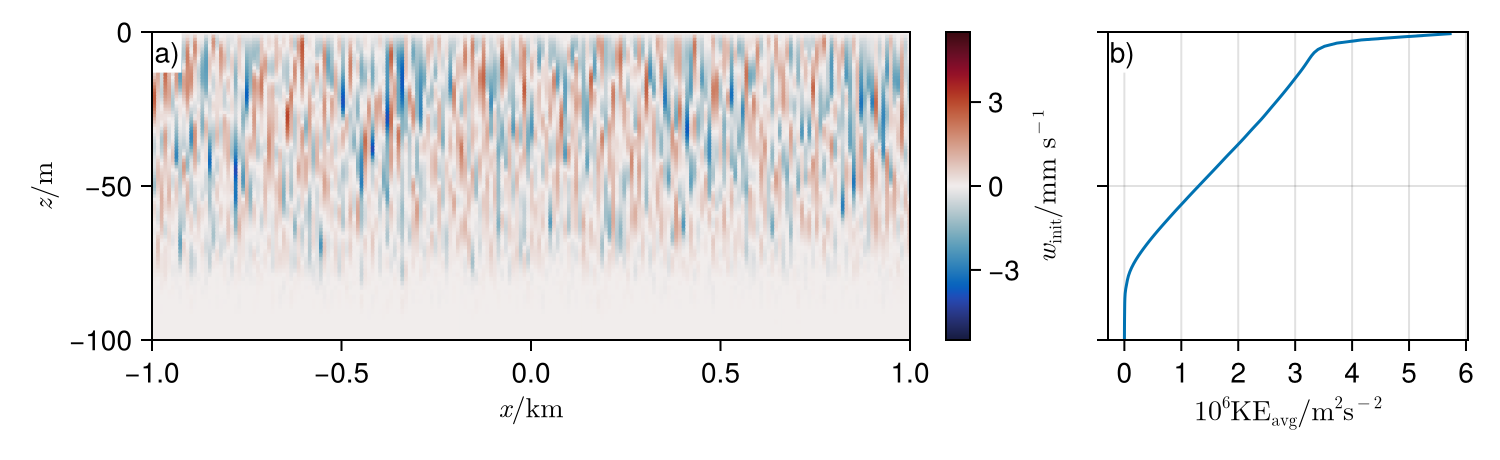}
    \caption{\Rthree{a) A slice of vertical velocity $w_\text{init}$ in the initial condition, just before the filament is imposed at $t=0$. b) Kinetic energy $\boldsymbol{u}_\text{init}\boldsymbol{\cdot}\boldsymbol{u}_\text{init} / 2$ at the same time, horizontally averaged, as a function of depth.}}
    \label{fig:noise}
\end{figure}

\Rthree{Before we impose the reference state, there is a brief initialisation process to ensure that the initial conditions, with noise, satisfy the Boussinesq equations to at least first order, as well as to allow the sub-grid parametrisation to reduce very large gradients caused by the pre-initial random noise before the unstable flow is introduced.} We pre-initialise the simulation at $ft/2\pi=-1$ with a state identical to the reference state at $|x|\rightarrow \infty$. We then add noise to all velocities and run the simulation until $t=0$. This produces fields $\boldsymbol u_{\text{init}} = \dfm{\boldsymbol {u}}_{\text{init}} + \boldsymbol {u}_{\text{init}}'$ and $b_{\text{init}} = \dfm{b}_{\text{init}} + b_{\text{init}}'$. \Rthree{The pre-initialised noise is sampled from a Gaussian distribution and is chosen to have similar magnitude and variation with depth of the turbulent kinetic energy to that which would be produced by forcing due to surface cooling of order $\sim 100\text{ Wm}^{-2}$ \citep[see][]{2018_JFM_SullivanMcWilliams}. A slice of vertical velocity and the horizontally-averaged kinetic energy is shown in figure \ref{fig:noise}. We explore the impact of the choice of initial noise on the conclusions of this study in appendix \ref{app:transient}} The initial state of the filament phase of the simulation is taken to be the reference state plus the fluctuation terms from the initialisation phase, namely,
\begin{equation*}
    \boldsymbol{u}(t=0) = v_0\hat{\boldsymbol{y}} + \boldsymbol{u}_{\text{init}}'\quad\text{and}\quad b(t=0) = b_0 + b_{\text{init}}'.
\end{equation*}

The mean, balanced state provides the energy source for instabilities with the nature of the instability determining the dominant energy source. \Rtwo{We consider contributions to the rate of change of the total (kinetic and potential) energy in the mean state, integrated over the whole domain $\mathcal{V}$. For the boundary conditions considered here, the rate of change of mean total energy may be partitioned into turbulent flux terms as follows,
\begin{equation}
\label{eq:energy-balance}
\deriv{}{t}\int_\mathcal{V} \text{d}V \left ( 
    \frac{1}{2}\dfm{\boldsymbol{u}} \boldsymbol{\cdot} \dfm{\boldsymbol{u}}
    - \dfm{b}z
\right ) 
\approx
\int_\mathcal{V}\text{d}V \bigg( 
    {\dfm{\boldsymbol{u}}_{,x} \boldsymbol{\cdot} \dfm{u'\boldsymbol{u}'}} 
    + {\dfm{\boldsymbol{u}}_{,z} \boldsymbol{\cdot} \dfm{w'\boldsymbol{u}'}} 
    - { \dfm{w'b'}}
\bigg ) ,
\end{equation}
wherein the direct impact of the sponge forcing and sub-grid dissipation on the reference state are assumed to be negligible.} Energisation of fluctuations by a laterally or vertically sheared velocity profile is given by the lateral or vertical shear production respectively
\begin{equation*}
    \text{LSP} = -\int_\mathcal{V} \text{d}V\; \dfm{\boldsymbol u}_{,x} \boldsymbol{\cdot} \dfm{u' \boldsymbol u'}, \quad \text{VSP} = -\int_\mathcal{V} \text{d}V\; \dfm{\boldsymbol u}_{,z} \boldsymbol{\cdot} \dfm{w' \boldsymbol u'}.
\end{equation*} 
\Rtwo{The VSP approximates the geostrophic shear production (GSP) if the down-front averaging can be assumed to separate the geostrophic and ageostrophic components. GSP provides the energy source for symmetric instability, among others. Lateral shear is known to energise inertial and barotropic instabilities.} Mean potential energy is transferred to fluctuations via the (vertical) buoyancy flux, \Rtwo{typically associated with baroclinic and gravitational instabilities},
\begin{equation*}
    \text{BFLUX} = \int_\mathcal{V} \text{d}V\; \dfm{w'b'}.
\end{equation*}

\section{Results}

\subsection{General evolution}
For the first 2--3 inertial periods \Rone{($\sim 54~\text{hr}$ at mid-latitudes)}, the filament remains approximately uniform in the down-front direction. Towards the end of our simulations \Rtwo{($ft/ 2\pi = 5.4$, not shown)}, and especially visible at the surface, the isopycnals begin to meander due to baroclinic instability. The domain is not large enough to support many baroclinic modes; following \citet{1966_JAS_Stone}, the fastest-growing baroclinic mode wavelength $\lambda$ can be estimated for an Eady model with parameters comparable to the centre of a jet,
\begin{equation}
    \lambda =\frac{2\pi v_0}{f}\sqrt{\frac{1+Ri_{\min{}}}{5/2}} \approx 5L,
    \quad 
    \label{eq:baroclinic-wavelength}
\end{equation}
which is half the downfront length of our domain. \Rtwo{The small domain limits the baroclinic growth rate. Nevertheless,} we ran a simulation with a longer down-front domain that resolved around eight wavelengths of this instability (not shown), \Rtwo{in which the meandering indeed becomes noticeable sooner, at around $ft/2\pi\approx 2$; \Roneb{and} verified that, even then, the baroclinic instability does not alter the conclusions we present in this study}. The rest of this discussion is concerned with the evolution of the filament before the meandering, which can be understood from the perspective of the down-front mean state. 

\begin{figure}
    \centering
    \includegraphics[width=\textwidth]{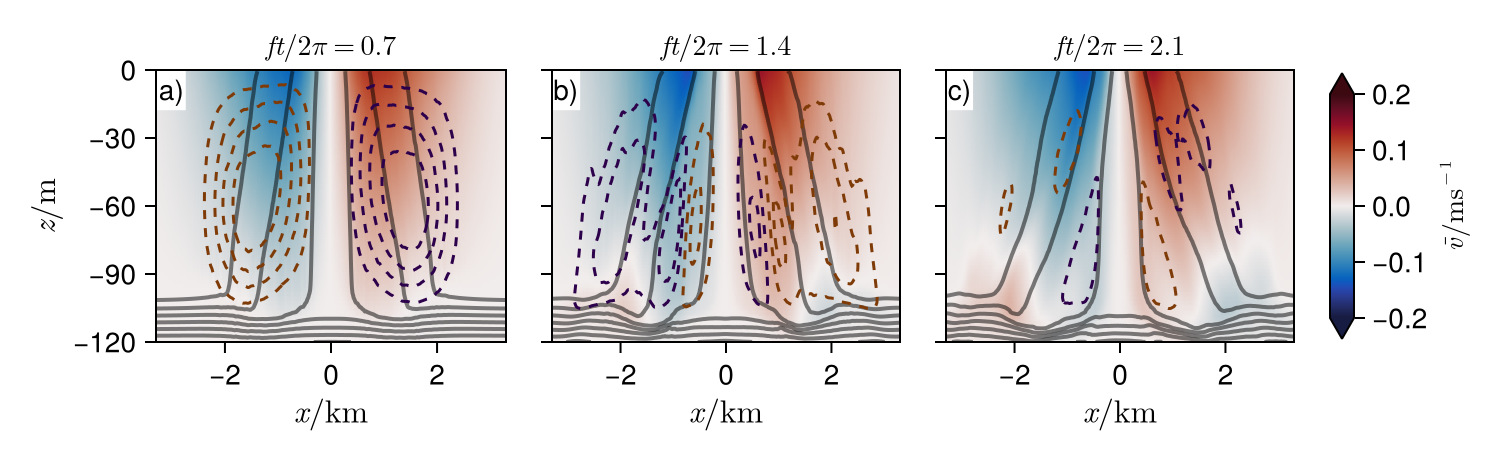}
    \caption{Evolution of the mean state for the simulation with $Ri_\text{min} = 0.0$. Three snapshots are shown at $ft/2\pi=\{0.8, 1.5, 2.2\}$. The heatmap shows $\dfm{v}$. \Rus{Solid} black contours are mean isopyncals $\dfm{b}$. \Rus{Dashed} orange (purple) contours are (anti-)clockwise streamlines \Rus{of the down-front mean flow}.
    }
    \label{fig:v-evolution}
\end{figure}

\begin{figure}
    \centering
    \includegraphics[width=\textwidth]{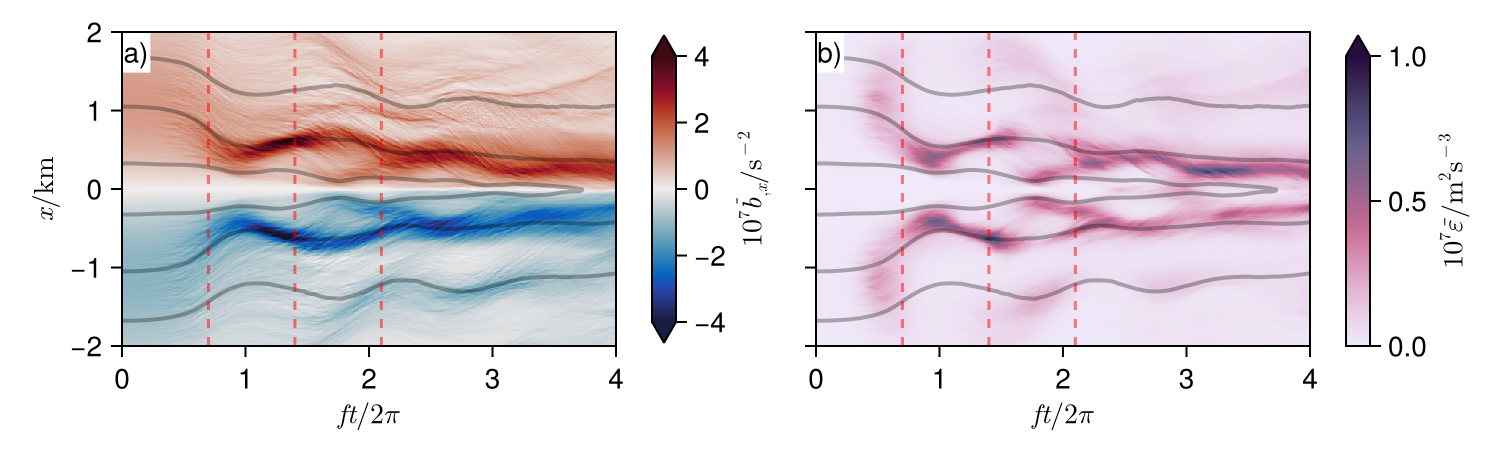}
    \caption{\Rthree{a) A Hovm\"oller plot of the horizontal buoyancy gradient of the down-front mean state for the simulation with $Ri_\text{min} = 0.0$, averaged over the top 10\% of the mixed layer. Black contours are contours of the buoyancy averaged over the same region. Red, dashed lines indicate the times at which the snapshots in figure \ref{fig:v-evolution} are taken. b) The down-front mean sub-grid dissipation of kinetic energy $\dfm{\varepsilon} = \dfm{\nu \nabla \boldsymbol{u}:\nabla \boldsymbol{u}}$, as in a).}}
    \label{fig:hovmoller}
\end{figure}

\Rone{The vertically sheared jet does not satisfy the no-flux boundary condition at $z=0$ for the down-front velocity and there is an onset of restratifying across-front Ekman flow; however by $ft/2\pi = 0.7$ this is overshadowed by a strong secondary circulation throughout the mixed layer, which is the focus of this section.}

Figure \ref{fig:v-evolution} outlines the evolution of the \Rone{down-front} mean state during the first inertial periods. Shortly after the filament is allowed to evolve freely, a strong secondary circulation on the scale of the fronts emerges (figure \ref{fig:v-evolution}a). The secondary circulation acts to restratify the fronts. At the surface, the circulation increases front strength: horizontal convergence pushes isopycnals together immediately under the surface. \Rthree{Afterwards, the secondary circulation changes sign (figure \ref{fig:v-evolution}b).} \Rthree{Hovm\"oller plots of the down-front-averaged horizontal buoyancy gradient and sub-grid dissipation, averaged over the down-front direction and the top 10\% of the mixed layer are shown in figure \ref{fig:hovmoller}. Near-surface isopycnals are drawn together, and horizontal buoyancy gradients reach a maximum strength at $ft/2\pi \approx 1.3$. This is associated with the maximum level of sub-grid dissipation. Afterwards, the down-front-averaged near-surface buoyancy gradients weaken before somewhat strengthening again, reaching a second maximum around one inertial period after the initial maximum. A sub-grid dissipation signal also recurs at this time ($ft/2\pi \approx 2.1$). The fronts approach a final state (disregarding the later meandering due to baroclinic instability)  with surface buoyancy gradients that are stronger than those in the reference state, though weaker than the maxima achieved earlier.}


\subsection{Initial evolution is a fast instability}

In this section, we show that the dynamics in the first inertial period contains an \Rus{instability that extracts energy from the the vertical sheared velocity}. 

\begin{figure}
    \centering
    \includegraphics[width=\textwidth]{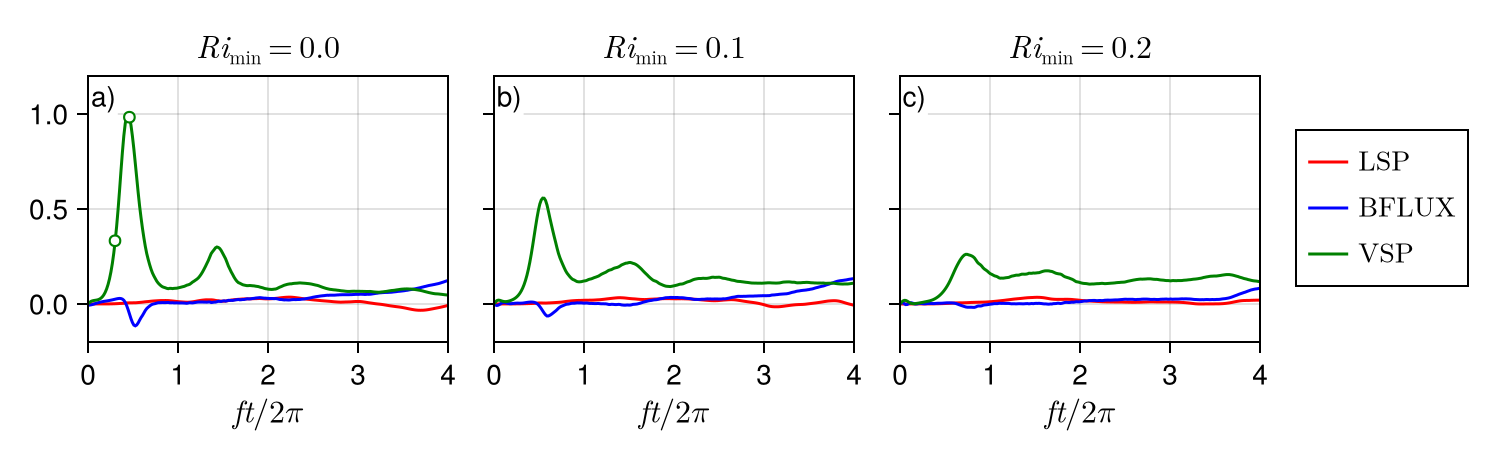}
    \caption{Terms extracting energy from the mean state for each of the simulations integrated over the mixed layer. Each term is normalised by the maximum VSP in the simulation with $Ri_{\min{}}=0.0$. Marked points in c) indicate the times at which the snapshots of vertical velocity and contribution to VSP in figures \ref{fig:wVSP1} and \ref{fig:wVSP2} are taken.}
    \label{fig:shear-production}
\end{figure}

\begin{figure}
    \centering
    \includegraphics[width=\textwidth]{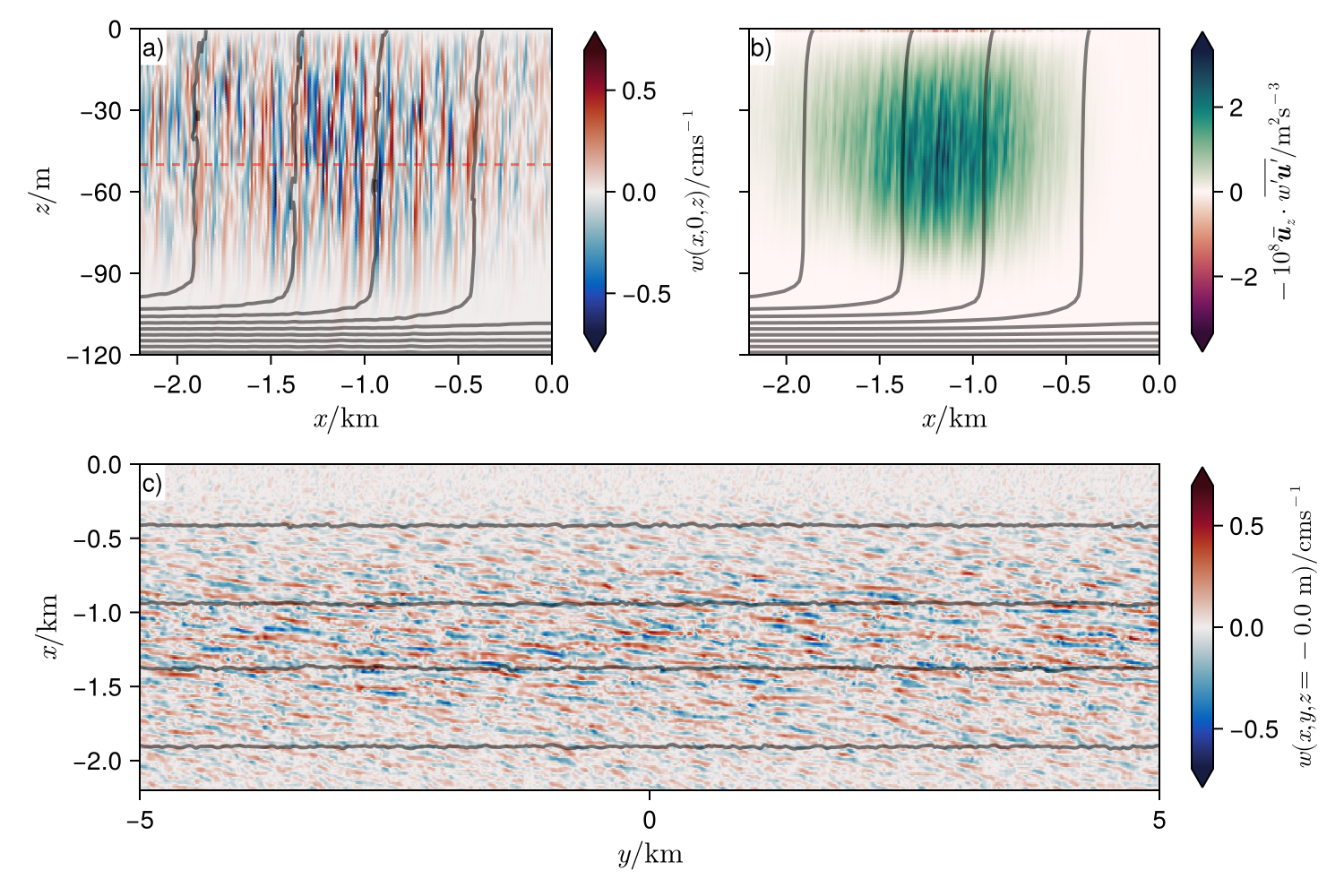}
    \caption{At $t/2\pi=0.30$, comparison of an $x$-$z$ slice at $y=0$ of a) vertical velocity $w$ and b) contribution to vertical shear production in the left side of the filament for the simulation with \Roneb{$Ri_\text{min}=0.0$}. The red, dashed line in the top left panel locates $z=-50\text{m}$, the depth of the $x$-$y$ slice of vertical velocity displayed in c). The horizontal slice of $w$ is coarse-grained with a Gaussian filter with a half-width of one grid cell for clarity. A video timeseries of this plot appears in the supplementary material as Movie 1.}
    \label{fig:wVSP1}
\end{figure}

\begin{figure}
    \centering
    \includegraphics[width=\textwidth]{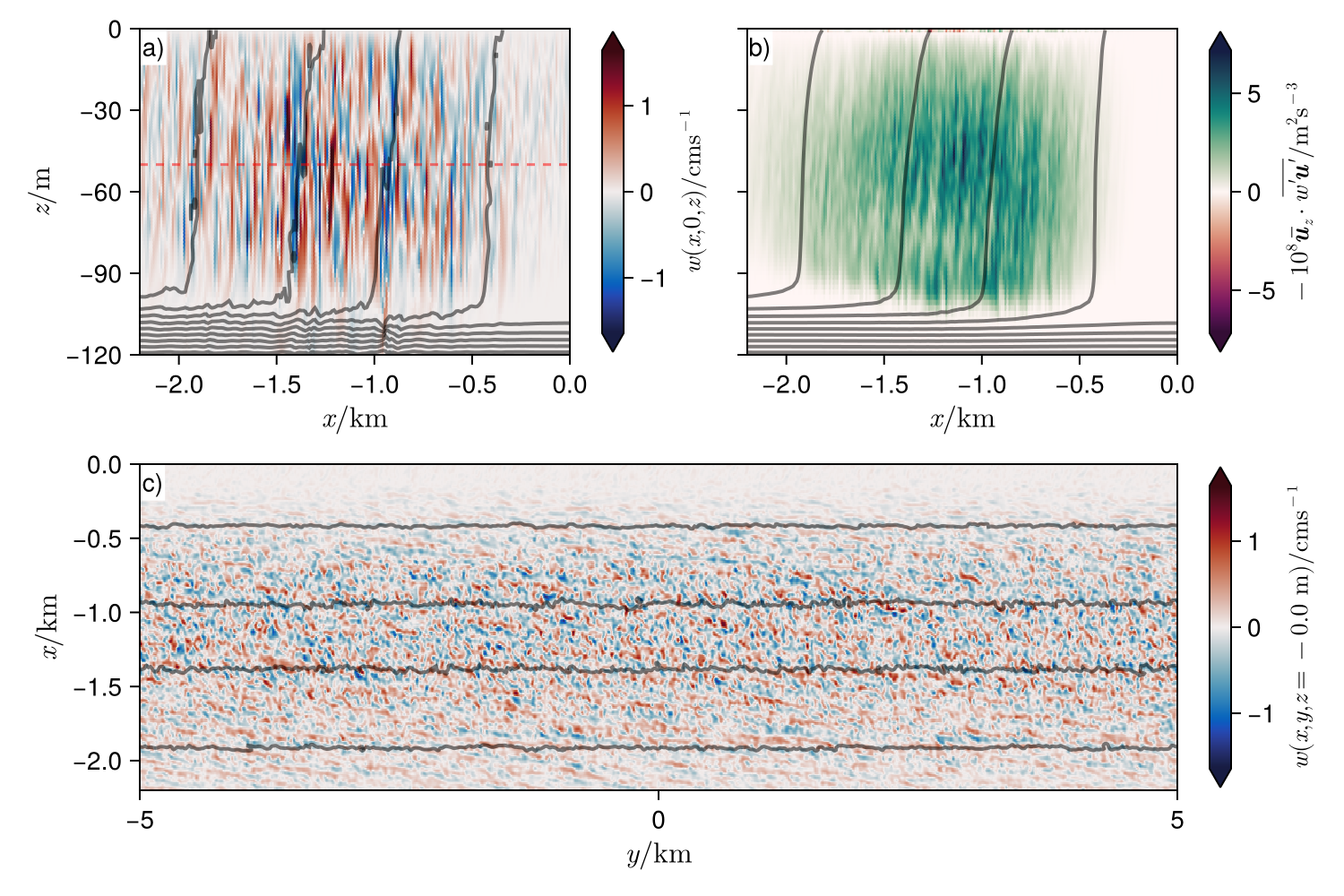}
    \caption{As in figure~\ref{fig:wVSP1} but at $t/2\pi = 0.46$, the time at which VSP is maximum (see \ref{fig:shear-production}). Note the difference in vertical velocity and shear production scales compared to figure \ref{fig:wVSP1}. }
    \label{fig:wVSP2}
\end{figure}

Figure \ref{fig:shear-production} shows an initial sharp peak in the VSP, greater for the simulations with smaller minimum Ri, before $t/2\pi \lessapprox 0.8$. Other sources of turbulent kinetic energy are negligible in comparison.

\Rthree{Figures \ref{fig:wVSP1} and \ref{fig:wVSP2} show the VSP and horizontal ($x$-$y$) and vertical ($x$-$z$) slices of the vertical velocity $w$ at $ft/2\pi=0.30$ and $0.46$ respectively for one of the fronts that comprises the filament.} At the earlier time, the contribution of the turbulence to vertical shear production is greater near the surface, \Rthree{and the horizontal slice shows some organisation within the front, though there is significant down-front variation.}

Once the instability saturates and gives way to turbulence, there is a large production of turbulent kinetic energy via VSP. The saturated regions in figure \ref{fig:wVSP1}a,b above $z=-50\text{ m}$ are associated with a larger VSP signature. The maximum in VSP moves down over time, \Rthree{indicating that the instability grows in a non-normal fashion,} and figure \ref{fig:wVSP2}c shows the horizontal state as the somewhat coherent cells are destroyed by secondary instabilities and the turbulence begins to isotropise in the $x$-$y$ plane. The extraction of energy is fast, and can be seen in figure \ref{fig:shear-production} to grow, peak and subside within the first inertial period. The consequence is that the filament is now out of geostrophic balance, as shown in the following section.

A video of the vertical velocity and mean contribution to VSP as shown in figures \ref{fig:wVSP1} and \ref{fig:wVSP2} is presented in the supplementary material. \Rtwo{Appendix \ref{app:transient} presents changes in behaviour as the large initial noise (figure \ref{fig:noise}) is lowered. Doing so recovers the symmetric normal modes as expected of the low-Ri initial conditions. While qualitatively distinct, the effect of the normal-mode SI on the filament is largely the same as the high-noise case, insofar as this study is concerned. That is, even though normal-mode SI grows somewhat more slowly than the non-normal instability we just described, it remains fast enough to cause a rapid imbalance of the jet, followed by a filament-scale inertial oscillation, similar to the one we are about to describe.}

\subsection{The initial instability echoes throughout the front as an inertial oscillation}

Following the onset of instability, the evolution of the filament is not balanced. Indeed, the process in the previous section happens within less than an inertial period, and in the resulting evolution the front does not stabilise at some minimum width. We highlight the consequences of the short-lived turbulent energy production using the linearised Sawyer-Eliassen equation for the mean secondary circulation streamfunction $\psi$ forced by by-products of turbulent vertical fluxes as we measure them in our simulations. That is, we use
\begin{equation}
    \nabla ^2 \pdel{\psi}{tt} + \underbrace{f^2 \pdel{\psi}{zz} - J(fv_0,\pdel{\psi}z)+ J(b_0, \pdel{\psi}x)}_{L} = \underbrace{f \pdel{\dfm{w'v'}}{zz} - \pdel{\dfm{w'b'}}{xz}}_{G},
    \label{eq:linearised-SE}
\end{equation}
in which we have neglected $O(\psi^2$) terms, and where
\refstepcounter{equation}
$$
  (\dfm{u}, 0, \dfm{v}) = (-\pdel{\psi}z, 0, \pdel{\psi}x), \quad \text{and}\quad
   J(a, b) = \pdel a z \pdel b x - \pdel b z \pdel a x.
  \eqno{(\theequation{\mathit{a},\mathit{b}})}\label{eq:psi-J}
$$
In doing so, we neglect terms due to the sub-grid momentum and buoyancy fluxes, as well as the remainder of the turbulent flux terms. The turbulent velocity flux $\dfm{w'v'}$ induces a forcing on $\psi$. Integrating this equation over the left half of the interior mixed layer gives a leading-order evolution equation for the circulation around the corresponding front $C_\text{IML} = -\iint \nabla^2\psi \di x\di y$, namely,
\begin{equation}
    \deriv{^2C_\text{IML}}{t^2} = \int_{-0.9H}^{-0.1H}\text{d}z \int^0_{-L_x/2} \text{d}x \Big [f^2 \pdel{\psi}{zz} - J(fv_0,\pdel{\psi}z)+ J(b_0, \pdel{\psi}x) - f \pdel{\dfm{w'v'}}{zz} + \pdel{\dfm{w'b'}}{xz} 
    \Big ].
    \label{eq:linearised-circulation}
\end{equation}

\begin{figure}
    \centering
    \includegraphics[width=\textwidth]{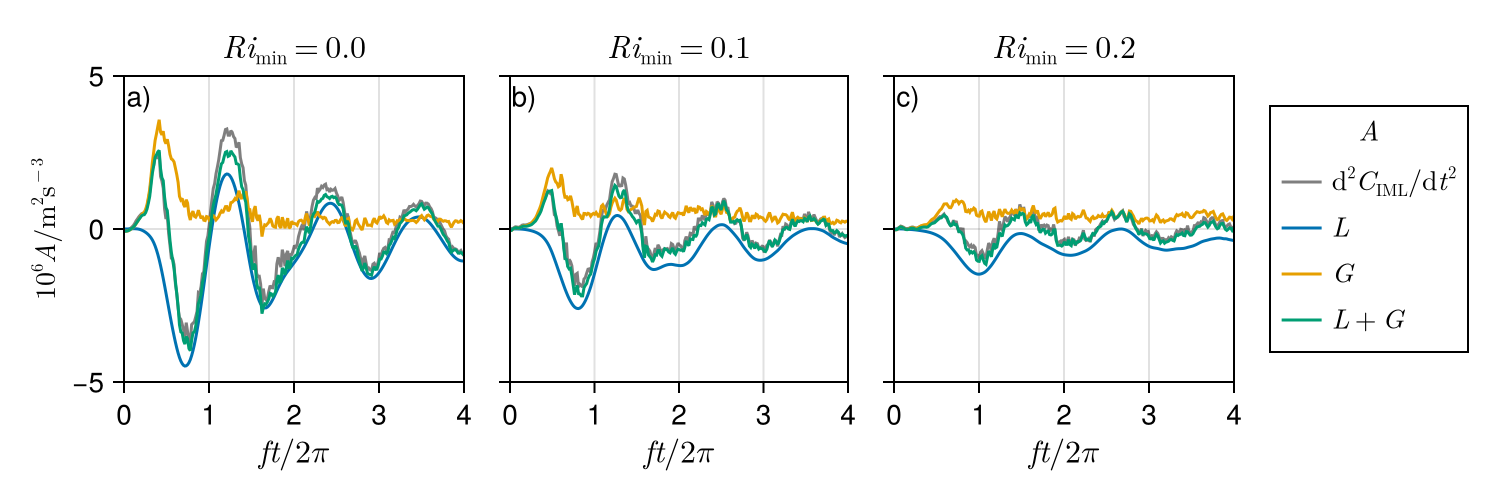}
    \caption{The circulation around the left half of the interior mixed layer as in equation \eqref{eq:linearised-circulation}. The black line is the actual evolution of the circulation; \Rus{i.e. the left side of \eqref{eq:linearised-circulation}}. The blue line is the contribution of the linear terms, the orange line is the contribution of the by-products of turbulent flux terms, and the green line is their sum. $L$ and $G$ are defined as in equation \eqref{eq:linearised-SE}. All timeseries are smoothed with a Gaussian kernel with standard deviation $0.1/f$.}
    \label{fig:streamfunction}
\end{figure}

The frontal circulation appears dominated by decaying near-inertial oscillations. The peak in \Rus{$G$} is large, and occurs in less than one inertial period, before significant change in the mean state; this \Rone{indicates} that the subsequent behaviour will be unbalanced and lead to oscillations. \Rone{To confirm, a} comparison of the terms in equation \eqref{eq:linearised-circulation} is shown for each of the simulations in figure \ref{fig:streamfunction}. The circulation shows clear decaying oscillatory behaviour with period $2\pi / f$, with linear terms dominating for small $Ri_{\min{}}$. The strength of these oscillations is dependent on the Richardson number of the flow: a smaller $Ri_{\min{}}$ produces a faster and stronger extraction of kinetic energy and in turn leads to higher amplitude oscillations.
\Rone{The secondary circulation is correlated with the surface front strength. The maximum front strength in figure \ref{fig:hovmoller}a emerges at $ft/2\pi \approx 1$, at the same time that isopycnals begin to move apart. This coincides with the moment \Rtwob{${\text{d}^2C}/{\text{d}t^2} =0$}, and the circulation changes sign (figure \ref{fig:streamfunction}a).}

\section{Discussion and conclusions}
We investigated the evolution of low-Richardson number mixed layer filaments with numerical simulations. \Rus{Energy is quickly extracted, via vertical shear production, from the down-front velocity}. This puts the filament in an unbalanced state, and results in vertically sheared inertial oscillations \Rone{that strengthen horizontal buoyancy gradients near the surface, while weakening them near the base of the mixed layer}. The oscillations are well-described by the linear terms in the Sawyer-Eliassen equation for the down-\Rus{front} streamfunction, following a forcing \Rus{induced by} the vertical turbulent flux of down-front momentum.


\Rtwo{The specific nature of the instability is outside the scope of this paper, but will no doubt be of interest to readers. In short, we believe it to be a non-normal, transient response towards SI that is sensitive to the initial noise, rather than normal-mode SI. While we leave a full demonstration to future studies, we find a comparison with reduced-noise simulations particularly illuminating. We present these results in appendix \ref{app:transient}. The initial noise illustrated in figure \ref{fig:noise} has a large magnitude, compared to the mean flow; and its vertical profile is responsible for the apparent downward propagation of the VSP contribution from figures \ref{fig:wVSP1} to \ref{fig:wVSP2}. At low levels of initial noise, the unstable modes that emerge are down-front symmetric (figure \ref{fig:m8}) before being destroyed by secondary instabilities (not shown). Perturbations emerge at the same rate throughout the mixed layer in a normal-mode fashion, rather than growing fastest at the surface. The low-noise behaviour is similar to the uniform-gradient Eady state studied by \citet{Wienkers2021}. The authors observe a similar instability-to-inertial-oscillations pathway that begins with a strong symmetric instability in 2.5D direct numerical simulations of finite-width, zero Richardson number frontal zones (i.e., constant lateral buoyancy gradients in domains that are periodic in the cross-front direction).} \Roneb{The generation of inertial oscillations and subsequent surface convergence and frontal strengthening is agnostic to the method of extraction of kinetic energy, as long as it is rapid enough.}

\Rthree{\citet{arobone2015effects}, \citet{zemskova2020transient}, and \citet{kimura2024initial} investigate the initial and long-time behaviour of a symmetrically unstable flow. The initial and normal growth of \Rthreeb{linear} perturbations in such a flow are different. In particular, the perturbations are initially strongly asymmetric and not aligned with isopycnals, before becoming nearly symmetric for large times \citep{arobone2015effects}. The instability at the start of the simulations in this paper is such an initial response of the filament associated with SI. \Rthreeb{Non-normal perturbations, seeded by the initial noise}, do not have time to grow into normal-mode SI before \Rthreeb{becoming significantly non-linear and} being isotropised by secondary instabilities acting on the growing modes \Rthreeb{and by interaction with the remainder of the initial noise}, leading to the rapid energy extraction from the mean state via the VSP.} \Rthreeb{The role and applicability of linear instability analyses in the real ocean, where internal waves, inertial waves and convection contribute to noise, which may strongly interact with both normal and transient forms of instability, is not clear and requires attention in future studies.}

Submesoscale structures in a realistic, time-varying ocean can show inertial oscillations associated with the diurnal variation of vertical diffusivity as shown by \citet{dauhajre2018diurnal}. We demonstrate here that the changes in effective diffusivity need not come from changes in surface forcing, but can arise without forcing, from turbulent kinetic energy production due to submesoscale instabilities acting on a front. \Rtwo{The resulting inertial oscillations contribute to surface frontal strengthening, but originate from unstable initial conditions, rather than a general feature of mixed layer frontogenesis in the real ocean, where instabilities, fronts, and turbulent mixing co-exist and evolve together.} \Rthree{With similarly unstable initial conditions, and the additional presence of a persistent turbulence-driven secondary circulation, we suspect that the inertial oscillations will be superimposed on a monotonic, turbulent thermal wind frontogenesis.}

Frontal features created in the prototypical sense, via a confluent strain flow, cross a range of scales before they reach the regime of the filament presented here. We believe that this work indicates there is understanding to be gained by spinning up fronts \textit{in-situ} for idealised numerical studies of turbulence, rather than beginning with what we think a submesoscale front ought to look like. In particular, the transition from $Ro \ll 1$ to $Ro \sim 1$ frontogenesis as these scales remain poorly resolved by global ocean models. \Rone{Of particular interest here, \citet{Thomas2012} showed that the behaviour of near-inertial oscillations is significantly altered in regions of frontogenesis driven by strain. Near-inertial waves are captured by growing fronts, which quickly rotate the wavevector to align with isopycnals, before kinetic energy extraction via the frontogenetic secondary circulation damps them completely. Therefore, we believe that in large-eddy simulations combining vertical convection and mixed layer strain-driven frontogenesis across scales, near-inertial oscillations of the nature described in this paper will not be present.}  

Submesoscale fronts and filaments are the sites of various instabilities, the consequences on the global ocean and best parametrisation of which remains an open question. Here we have shown that instability alone is capable of strengthening surface buoyancy gradients in a mixed layer filament via a multi-stage process in which the initial instability, \Rthree{which is likely a non-normal, transient response to initial noise}, acts quickly and puts the fronts out of balance. The resulting adjustment begins with a convergent secondary circulation on the scale of the fronts that strengthens them near the surface. For our experiments, \Rone{arrest of surface convergence} is not \Rus{exclusively} due to turbulence and is \Rus{at least in part} the restoring phase of the inertial oscillation. We also comment on the consequences for studies of submesoscale flows, and suggest that the most fruitful way to investigate how these structures evolve in the real ocean, from a computational perspective, \Rthree{may require spinning up --- rather than imposing --- initial conditions, e.g. starting with a larger scale mixed layer front and applying an advective forcing,} to avoid energising unwanted dynamics due to instability or otherwise.

\backsection[Acknowledgements]{The authors ackowledge fruitful discussions with William R.\ Young, Leif N.\ Thomas, John Taylor and Matthew N.\ Crowe, \Rusb{as well as three anonymous reviewers for their illuminating critique and suggestions}. Computations were performed on the Mist supercomputer at the SciNet HPC Consortium \citep{loken2010scinet, ponce2019deploying}. SciNet is funded by Innovation, Science and Economic Development Canada; the Digital Research Alliance of Canada; the Ontario Research Fund: Research Excellence; and the University of Toronto.}
\backsection[Funding]{E.A.\ and N.G.\ acknowledge the support of the Natural Sciences and Engineering Research Council of Canada (NSERC) [funding reference number RGPIN-2022-04560].}
\backsection[Declaration of interests]{The authors report no conflict of interest.}
\backsection[Data availability statement]{Replication code for this manuscript can be found at \url{https://doi.org/10.5281/zenodo.14852129} \citep{atkinson_2025_14852129}}
\backsection[Author ORCIDs]{E.\ Atkinson, https://orcid.org/0000-0002-5913-3137; J.C.\ McWilliams, https://orcid.org/0000-0002-1237-5008; N.\ Grisouard, https://orcid.org/0000-0003-4045-2143}

\appendix
\section{Parametrisation of the reference state}
\label{app:filament}
The reference state of the filament is assumed to be independent of the down-filament direction and described by a buoyancy profile
\begin{equation}
    b_0(x, z) = N_0^2 z + (N^2 - N_0^2)\lambda H\; g\!\left(\frac{z - h(x)}{\lambda H}\right),
    \label{eq:refbuoyancy}
\end{equation}
with a jet $v_0(x, z)$ in thermal wind balance ($f {\partial v_0}/{\partial z} = {\partial b_0}/{\partial x}$), with $v_0\rightarrow 0$ as $z \rightarrow - \infty$. The above equations are defined in terms of a filament shape function 
\refstepcounter{equation}
$$
  h(x) = H\gamma\left(\frac{x}{\ell}\right)\quad\text{with}\quad
  \gamma(s) = -1 + \frac{\delta}{2} \left [ \text{erf} \left ( s + \frac{1}{2\alpha}\right ) - \text{erf} \left ( s - \frac{1}{2\alpha}\right )\right ].
  \eqno{(\theequation{\mathit{a},\mathit{b}})}\label{eq:h}
$$
and mixed layer transition function $g(\xi) = \ln (1 + e^{\xi})$. $N_0$ and $N$ are deep-water and mixed layer buoyancy frequencies respectively, $f$ is the Coriolis frequency, $L$ is the distance between front centres, $H$ is the depth of the mixed layer and $\ell$ is the frontal half-width. \Roneb{Remaining, non-dimensional, parameters are described in table \ref{tab:parameters}.}
We parametrise the state by varying $N$ and $N_0$ to achieve prescribed minimum local Rossby and Richardson numbers, given respectively by 
\refstepcounter{equation}
$$
  Ro_\text{min} = \frac{(N^2 - N_0^2)\beta^2}{f^2\alpha^2} \max_s \Big (\gamma \gamma _{,s}\Big )\quad\text{and}\quad
  Ri_\text{min} = \frac{f^2 \alpha^2 N^2}{\beta^2 (N^2 - N_0^2)^2} \min_s \Big ( \gamma_{,s}^{-2}\Big ).
  \eqno{(\theequation{\mathit{a},\mathit{b}})}\label{eq:RoRi}
$$
\Roneb{Target Rossby and Richardson numbers are given in table \ref{tab:parameters}.}
\begin{table}
    \centering
    \begin{tabular}{ccc}
        Front width-separation ratio & $\alpha = \ell/L $ & 1.5\\
        Mixed layer aspect ratio & $\beta = H/L$ & 0.1 \\
        Mixed layer depth change & $\delta$ & -0.25\\
        Mixed layer transition width& $\lambda$ & 0.025\\
        Minimum Rossby number & $Ro_\text{min}$ & -0.8 \\
        Minimum Richardson number & $Ri_\text{min}$ & $\in \{0, 0.1, 0.2\}$
    \end{tabular}
    \caption{Filament parameters}
    \label{tab:parameters}
\end{table}


\section{\Rthree{Transient behaviour and the impact of initial noise}}
\label{app:transient}
In the main body of this paper, we outline the process leading to inertial oscillations in an unstable front. In this additional section, we will attempt to connect the topic of this study with symmetric instability literature. We find that the emergence and propagation of the initial disturbance throughout the filament is dependent on the amplitude and form of the fluctuations in the initial conditions. This is beyond the scope of the main body in this paper, which is concerned with the consequences of unstable initial conditions for frontogenesis studies, but we include this secondary set of simulations and analyses to provide a more thorough exploration of the generation of inertial oscillations, as well as to demonstrate that the conclusions in the main body of the text hold for rather different circumstances.

Due to the choice of pre-initial conditions, the initialisation described in section \ref{sec:simulations} produces a state with a turbulent kinetic energy profile that is approximately linear in $z$ (from zero at the base of the mixed layer to a maximum near the thermocline). This serves as an analogue for the TKE of a state driven by surface cooling. This profile is the cause of the downward propagation of the VSP from figure \ref{fig:wVSP1} to \ref{fig:wVSP2}. Fluctuations simply start larger at the surface. Altering the noise profile to be constant in depth in the mixed layer, for example, causes the VSP to emerge approximately at the same time throughout the unstable region (not shown).

Here we present a set of three simulations, each with different values for the maximum (surface) amplitude of the pre-initial noise. The simulation in the main body of this paper has maximum pre-initial noise amplitude a factor $10^{-2}$ of the maximum down-front velocity in the reference state. In this section we repeat the analysis in the main body of the paper, but for simulations pre-initialised with noise of relative magnitude $10^{-4}$, $10^{-6}$, $10^{-8}$.

\begin{figure}
    \centering
    \includegraphics[width=\textwidth]{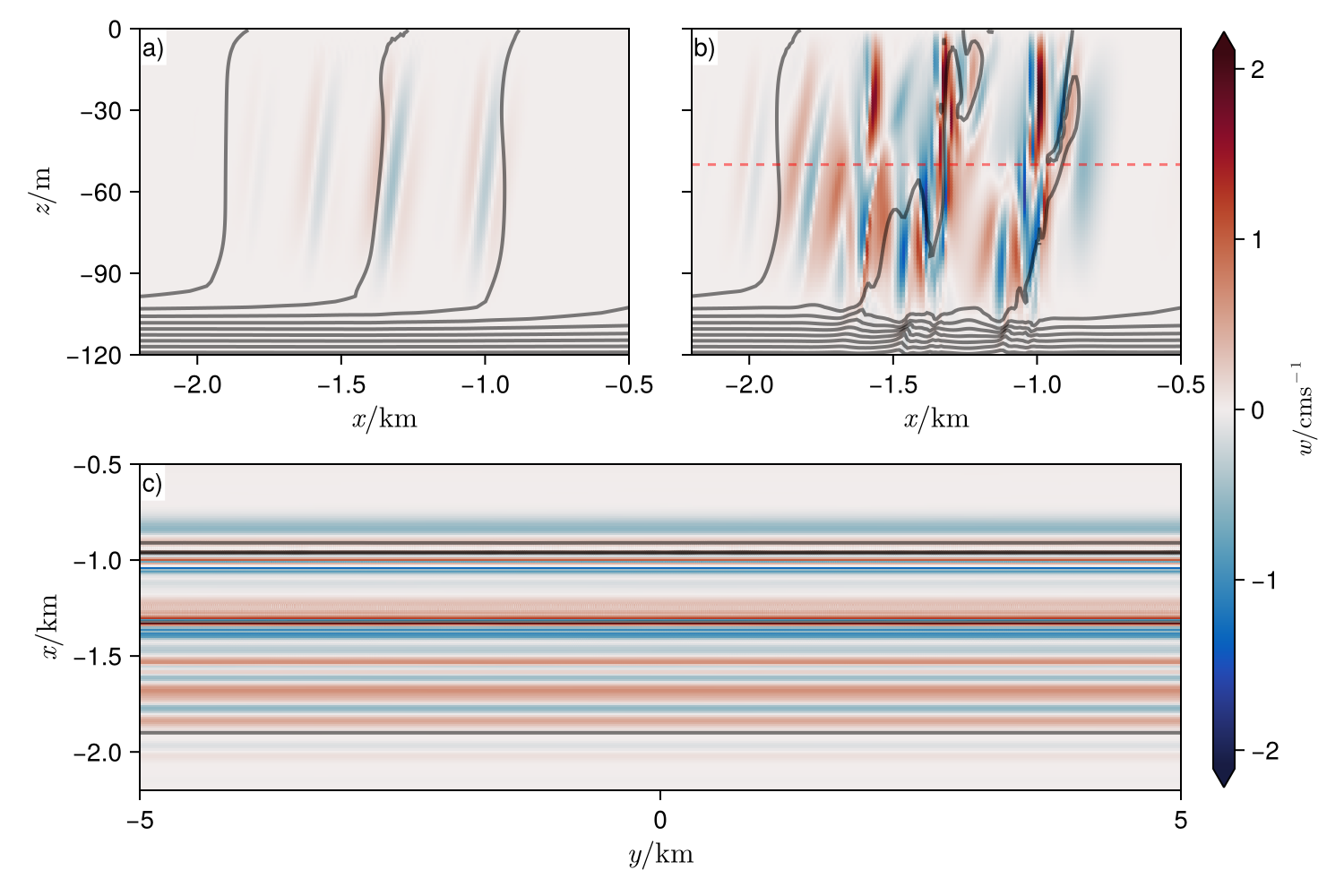}
    \caption{a, b) vertical slices of vertical velocity of the simulation with smallest pre-initial noise amplitude ($10^{-8}$) at a) $ft/2\pi = 0.8$ and b) $ft/2\pi = 1.0$ the time of maximum $\text{VSP}_0$ (see figure \ref{fig:amplitude-tke}). The red, dashed line in the top right panel locates $z=-50\text{ m}$, the depth of the $x$-$y$ slice of vertical velocity displayed in c). c) a horizontal slice of vertical velocity at $z=-50\text{ m}$ at the time of maximum $\text{VSP}_0$.}
    \label{fig:m8}
\end{figure}

For the simulation with smallest pre-initial noise amplitude, the form of the vertical velocity at early times is shown in figure \ref{fig:m8}, \Rusb{and a video timeseries of vertical velocity slices and VSP (as in figures \ref{fig:wVSP1} and \ref{fig:wVSP2}) appears in the supplementary material as Movie 2}. Characteristic SI modes are recovered. These modes are somewhat tilted with respect to the isopycnals \Rusb{in the x-z plane}, expected for non-hydrostatic SI \citep{Wienkers2021}.

The down-front mean defined in the earlier sections does not remove the large across-front wavenumber symmetric modes and as such is not appropriate nor illustrative to describe a coarse-grained state. With this in mind, and for this section only, we define the vertical and lateral shear productions, as well as the vertical buoyancy flux, using the departure of the simulation state from the reference filament state, namely,
\begin{equation}
    \label{eq:LSPVSLP0}
    \text{LSP}_0 = -\int_\mathcal{V} \text{d}V\; \ \frac{\partial \boldsymbol{u}_0}{\partial x}\boldsymbol{\cdot} u (\boldsymbol{u} - \boldsymbol{u}_0), \quad \text{VSP}_0 = -\int_\mathcal{V} \text{d}V\; \ \frac{\partial \boldsymbol{u}_0}{\partial z}\boldsymbol{\cdot} w (\boldsymbol{u} - \boldsymbol{u}_0), 
\end{equation} 
and 
\begin{equation}
    \label{eq:BFLUX0}
    \text{BFLUX}_0 = \int_\mathcal{V} \text{d}V\; w(b - b_0).
\end{equation}

We continue with these alternate definitions, noting that the perturbations from the reference state will include the down-front-mean ageostrophic circulation and may only be appropriate at early times. Nevertheless, figure \ref{fig:shear-production} was reproduced for the first set of simulations with the above definitions (not shown) and bears only quantitative differences.

\begin{figure}
    \centering
    \includegraphics[width=\textwidth]{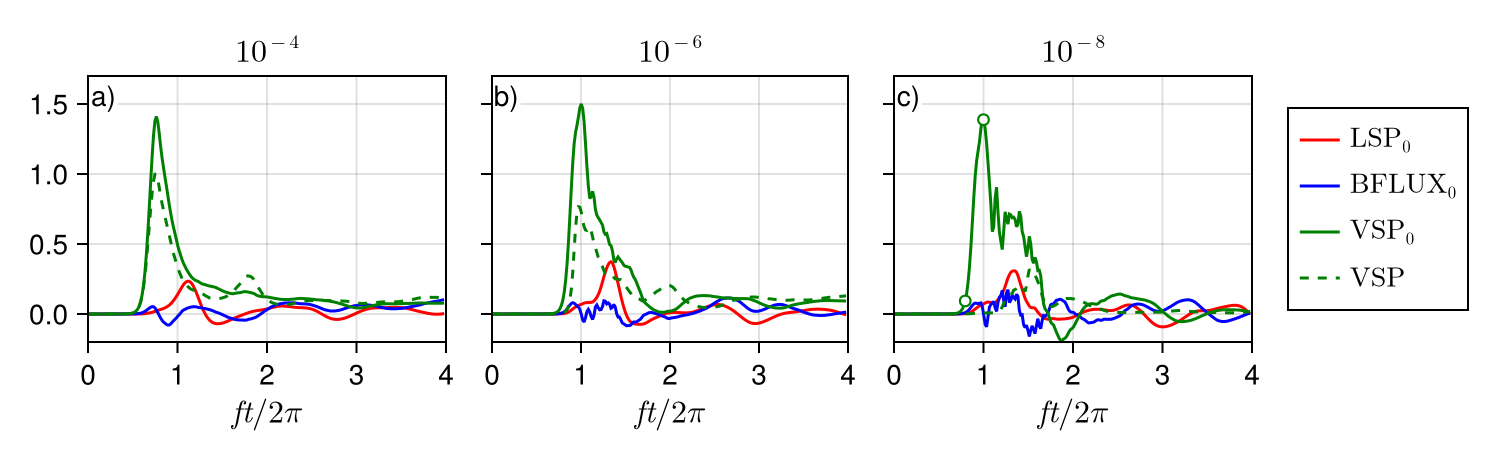}
    \caption{As in figure \ref{fig:shear-production} but for a series of simulations with decreasing amplitudes of initial noise relative to the maximum velocity, and using the TKE production definitions in equations \eqref{eq:LSPVSLP0} and \eqref{eq:BFLUX0}. From left to right, the simulations are pre-initialised with noise of relative magnitudes $10^{-4}$, $10^{-6}$, and $10^{-8}$. Terms are shown relative to the same scale in figure \ref{fig:shear-production}. Additionally, and for comparison, the VSP calculated according to equation \eqref{eq:energy-balance} is included as a dashed green line. Marked points in c) indicate the times at which the snapshots of vertical velocity in figure \ref{fig:m8} are taken.}
    \label{fig:amplitude-tke}
\end{figure}

Figure \ref{fig:amplitude-tke} shows the shear production and buoyancy flux terms for a set of simulations with decreasing amplitude of initial noise. In addition, it includes the VSP as defined in equation \eqref{eq:energy-balance}. The $\text{VSP}_0$ is the primary source of energy for the fluctuations, and behaviour across all cases up to the extremum in $\text{VSP}_0$ is largely identical; though, for the largest two noise amplitudes ($10^{-2}$, in figure \ref{fig:shear-production} and $10^{-4}$), the $\text{VSP}_0$ growth and peak occurs earlier. As expected, VSP associated with the perturbations vanishes as they become more symmetric at lower noise levels.

\begin{figure}
    \centering
    \includegraphics[width=\textwidth]{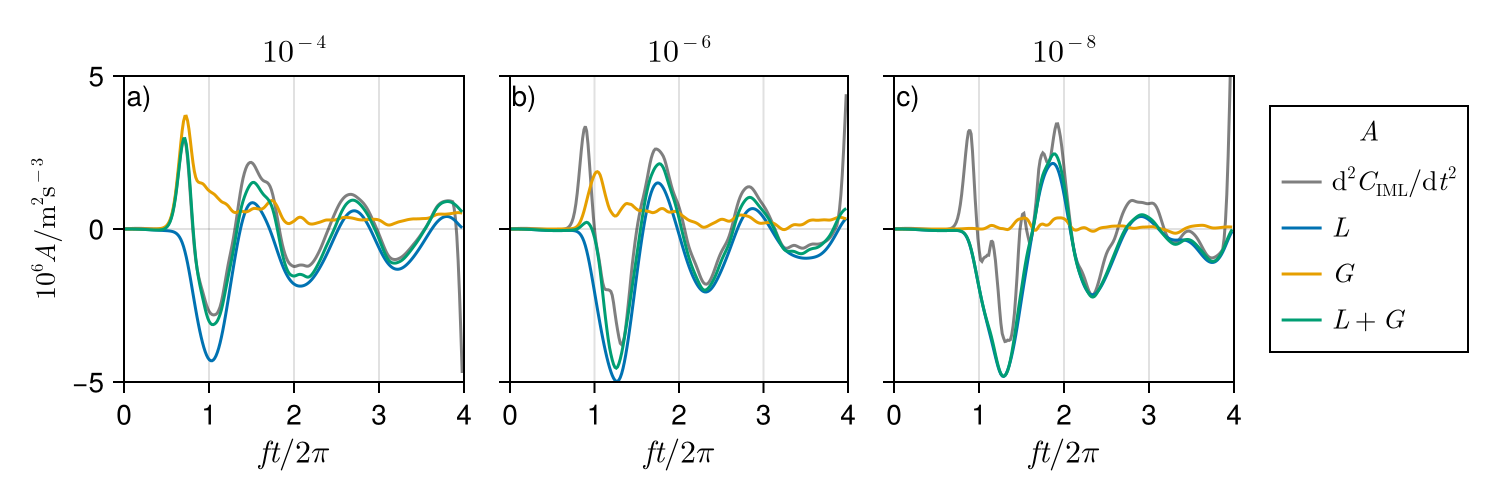}
    \caption{As in figure \ref{fig:streamfunction} but for a series of simulations with decreasing amplitudes of initial noise relative to the maximum velocity. From left to right, the simulations are pre-initialised with noise of relative magnitude $10^{-4}$, $10^{-6}$, and $10^{-8}$. All timeseries are smoothed with a Gaussian kernel with standard deviation $0.3/f$.}
    \label{fig:amplitude-streamfunction}
\end{figure}

\begin{figure}
    \centering
    \includegraphics[width=\textwidth]{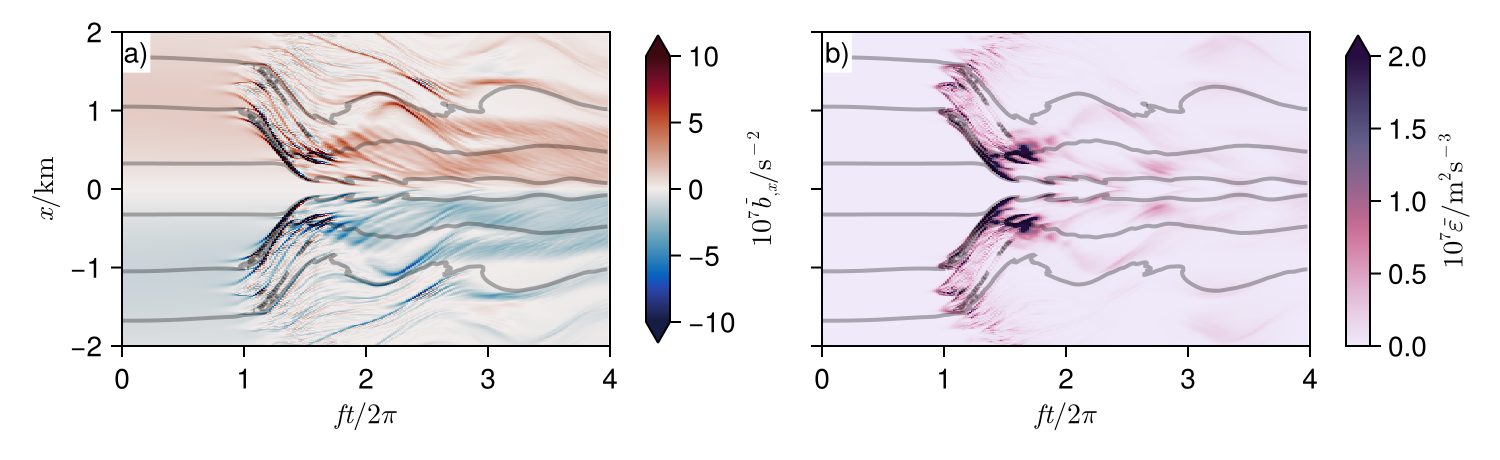}
    \caption{a) A Hovm\"oller plot of the horizontal buoyancy gradient of the down-front mean state for the simulation with smallest pre-initial noise amplitude ($10^{-8}$), averaged over the top 10\% of the mixed layer. Black contours are contours of the buoyancy averaged over the same region. b) The down-front mean sub-grid dissipation of kinetic energy $\dfm{\varepsilon} = \dfm{\nu \nabla \boldsymbol{u}:\nabla \boldsymbol{u}}$, as in a).}
    \label{fig:hovmoller-m8}
\end{figure}

Following the TKE production by $\text{VSP}_0$, inertial oscillations are also observed in the secondary circulations, as seen in figure \ref{fig:amplitude-streamfunction}. Hovm\"oller plots in figure \ref{fig:hovmoller-m8} show that the near-surface isopycnals follow the inertial oscillations, with a large front strength and enhanced dissipation just after the filament reaches a minimum width ($ft / 2\pi \approx 1.7$). 

\bibliographystyle{jfm}
\bibliography{bibliography}

\begin{thebibliography}{44}
\expandafter\ifx\csname natexlab\endcsname\relax\def\natexlab#1{#1}\fi
\def\au#1{#1} \def\ed#1{#1} \def\yr#1{#1}\def\at#1{#1}\def\jt#1{\textit{#1}} \def\bt#1{#1}\def\bvol#1{\textbf{#1}} \def\vol#1{#1} \def\pg#1{#1} \def\publ#1{#1}\def\arxiv#1{#1}\def\org#1{#1}\def\st#1{\textit{#1}}

\bibitem[Aravind {\em et~al.\/}(2023)Aravind, Verma, Sarkar, Freilich, Mahadevan, Haley, Lermusiaux \& Allshouse]{Aravind2023}
{\sc \au{Aravind, H.~M.}, \au{Verma, Vicky}, \au{Sarkar, Sutanu}, \au{Freilich, Mara~A.}, \au{Mahadevan, Amala}, \au{Haley, Patrick~J.}, \au{Lermusiaux, Pierre~F.J.} \& \au{Allshouse, Michael~R.}} \yr{2023}  \at{Lagrangian surface signatures reveal upper-ocean vertical displacement conduits near oceanic density fronts}.  \jt{Ocean Modelling}  \bvol{181}.

\bibitem[Arobone \& Sarkar(2015)]{arobone2015effects}
{\sc \au{Arobone, Eric} \& \au{Sarkar, Sutanu}} \yr{2015}  \at{Effects of three-dimensionality on instability and turbulence in a frontal zone}.  \jt{Journal of Fluid Mechanics}  \bvol{784},  \pg{252--273}.

\bibitem[Atkinson(2025)]{atkinson_2025_14852129}
{\sc \au{Atkinson, Erin}} \yr{2025} Filament instability and near inertial echoes - setup and analysis.

\bibitem[Barkan {\em et~al.\/}(2019)Barkan, Molemaker, Srinivasan, McWilliams \& D’Asaro]{2019_JPO_BarkanMSMD}
{\sc \au{Barkan, Roy}, \au{Molemaker, M.~Jeroen}, \au{Srinivasan, Kaushik}, \au{McWilliams, James~C.} \& \au{D’Asaro, Eric~A.}} \yr{2019}  \at{The role of horizontal divergence in submesoscale frontogenesis}.  \jt{Journal of Physical Oceanography}  \bvol{49},  \pg{1593--1618}.

\bibitem[Bodner {\em et~al.\/}(2023)Bodner, Fox-Kemper, Johnson, Roekel, Mcwilliams, Sullivan, Hall \& Dong]{Bodner2023}
{\sc \au{Bodner, Abigail~S}, \au{Fox-Kemper, Baylor}, \au{Johnson, Leah}, \au{Roekel, Luke P~Van}, \au{Mcwilliams, James~C}, \au{Sullivan, Peter~P}, \au{Hall, Paul~S} \& \au{Dong, Jihai}} \yr{2023}  \at{Modifying the mixed layer eddy parameterization to include frontogenesis arrest by boundary layer turbulence}.  \jt{Journal of Physical Oceanography}  \bvol{53},  \pg{323--339}.

\bibitem[Bosse {\em et~al.\/}(2021)Bosse, Testor, Damien, Estournel, Marsaleix, Mortier, Prieur \& Taillandier]{Bosse2021}
{\sc \au{Bosse, Anthony}, \au{Testor, Pierre}, \au{Damien, Pierre}, \au{Estournel, Claude}, \au{Marsaleix, Patrick}, \au{Mortier, Laurent}, \au{Prieur, Louis} \& \au{Taillandier, Vincent}} \yr{2021}  \at{Wind-forced submesoscale symmetric instability around deep convection in the northwestern mediterranean sea}.  \jt{Fluids}  \bvol{6}.

\bibitem[Brannigan(2016)]{Brannigan2016}
{\sc \au{Brannigan, L.}} \yr{2016}  \at{Intense submesoscale upwelling in anticyclonic eddies}.  \jt{Geophysical Research Letters}  \bvol{43},  \pg{3360--3369}.

\bibitem[Calil {\em et~al.\/}(2021)Calil, Suzuki, Baschek \& da~Silveira]{Calil2021}
{\sc \au{Calil, Paulo~H.R.}, \au{Suzuki, Nobuhiro}, \au{Baschek, Burkard} \& \au{da~Silveira, Ilson~C.A.}} \yr{2021}  \at{Filaments, fronts and eddies in the cabo frio coastal upwelling system, brazil}.  \jt{Fluids}  \bvol{6}.

\bibitem[Crowe \& Taylor(2018)]{2018_JFM_CroweTaylor}
{\sc \au{Crowe, Matthew~N.} \& \au{Taylor, John~R.}} \yr{2018}  \at{The evolution of a front in turbulent thermal wind balance. part 1. theory}.  \jt{Journal of Fluid Mechanics}  \bvol{850},  \pg{179--211}.

\bibitem[Dauhajre \& McWilliams(2018)]{dauhajre2018diurnal}
{\sc \au{Dauhajre, Daniel~P} \& \au{McWilliams, James~C}} \yr{2018}  \at{Diurnal evolution of submesoscale front and filament circulations}.  \jt{Journal of Physical Oceanography}  \bvol{48}~(10),  \pg{2343--2361}.

\bibitem[Emanuel(1979)]{Emanuel1979}
{\sc \au{Emanuel, Kerry~A}} \yr{1979}  \at{Inertial instability and mesoscale convective systems. part i: Linear theory of inertial instability in rotating viscous fluids}.  \jt{Journal of Atmospheric Sciences}  \bvol{36},  \pg{2425--2449}.

\bibitem[Freilich \& Mahadevan(2021)]{Freilich2021}
{\sc \au{Freilich, Mara} \& \au{Mahadevan, Amala}} \yr{2021}  \at{Coherent pathways for subduction from the surface mixed layer at ocean fronts}.  \jt{Journal of Geophysical Research: Oceans}  \bvol{126}.

\bibitem[Gula {\em et~al.\/}(2014)Gula, Molemaker \& McWilliams]{2014_JPO_GulaMM}
{\sc \au{Gula, Jonathan}, \au{Molemaker, M.~Jeroen} \& \au{McWilliams, James~C.}} \yr{2014}  \at{Submesoscale cold filaments in the gulf stream}.  \jt{Journal of Physical Oceanography}  \pg{p. 140728145042006}.

\bibitem[Gula {\em et~al.\/}(2022)Gula, Taylor, Shcherbina \& Mahadevan]{gula2022submesoscale}
{\sc \au{Gula, Jonathan}, \au{Taylor, John}, \au{Shcherbina, Andrey} \& \au{Mahadevan, Amala}} \yr{2022}  \at{Submesoscale processes and mixing}.  \bt{In {\em Ocean mixing\/}},  \pg{pp. 181--214}.  \publ{Elsevier}.

\bibitem[Haine \& Marshall(1998)]{1998_JPO_HaineMarshall}
{\sc \au{Haine, Thomas W.~N.} \& \au{Marshall, John}} \yr{1998}  \at{Gravitational, symmetric, and baroclinic instability of the ocean mixed layer}.  \jt{Journal of Physical Oceanography}  \bvol{28},  \pg{634--658}.

\bibitem[Harris {\em et~al.\/}(2022)Harris, Poulin \& Lamb]{Harris2022}
{\sc \au{Harris, M.~W.}, \au{Poulin, F.~J.} \& \au{Lamb, K.~G.}} \yr{2022}  \at{Inertial instabilities of stratified jets: Linear stability theory}.  \jt{Physics of Fluids}  \bvol{34}.

\bibitem[Heifetz \& Farrell(2007)]{2007_JAS_HeifetzFarrell}
{\sc \au{Heifetz, Eyal} \& \au{Farrell, Brian~F.}} \yr{2007}  \at{Generalized stability of nongeostrophic baroclinic shear flow. part ii: Intermediate richardson number regime}.  \jt{Journal of the Atmospheric Sciences}  \bvol{64},  \pg{4366--4382}.

\bibitem[Hoskins \& Bretherton(1972)]{Hoskins1972}
{\sc \au{Hoskins, B.J.} \& \au{Bretherton, F.P.}} \yr{1972}  \at{Atmospheric frontogenesis models: Mathematical formulation and solution}.  \jt{Journal of the Atmospheric Sciences}  \pg{pp. 11--37}.

\bibitem[Kimura(2024)]{kimura2024initial}
{\sc \au{Kimura, Satoshi}} \yr{2024}  \at{Initial and transient growth of symmetric instability}.  \jt{Journal of Physical Oceanography}  \bvol{54}~(1),  \pg{115--129}.

\bibitem[Klein {\em et~al.\/}(2019)Klein, Lapeyre, Siegelman, Qiu, Fu, Torres, Su, Menemenlis \& Gentil]{Klein2019}
{\sc \au{Klein, Patrice}, \au{Lapeyre, Guillaume}, \au{Siegelman, Lia}, \au{Qiu, Bo}, \au{Fu, Lee~Lueng}, \au{Torres, Hector}, \au{Su, Zhan}, \au{Menemenlis, Dimitris} \& \au{Gentil, Sylvie~Le}} \yr{2019}  \at{Ocean-scale interactions from space}.  \jt{Earth and Space Science}  \bvol{6},  \pg{795--817}.

\bibitem[Lilly(1967)]{lilly1967representation}
{\sc \au{Lilly, Douglas~K}} \yr{1967} The representation of small-scale turbulence in numerical simulation experiments.  \bt{In {\em Proc. IBM Sci. Comput. Symp. on Environmental Science\/}},  \pg{pp. 195--210}.

\bibitem[Loken {\em et~al.\/}(2010)Loken, Gruner, Groer, Peltier, Bunn, Craig, Henriques, Dempsey, Yu, Chen {\em et~al.\/}]{loken2010scinet}
{\sc \au{Loken, Chris}, \au{Gruner, Daniel}, \au{Groer, Leslie}, \au{Peltier, Richard}, \au{Bunn, Neil}, \au{Craig, Michael}, \au{Henriques, Teresa}, \au{Dempsey, Jillian}, \au{Yu, Ching-Hsing}, \au{Chen, Joseph} \& \au{others}} \yr{2010} Scinet: lessons learned from building a power-efficient top-20 system and data centre.  \bt{In {\em Journal of physics: conference series\/}}, ,  \vol{vol. 256},  \pg{p. 012026}. IOP Publishing.

\bibitem[Mahadevan(2016)]{Mahadevan2016}
{\sc \au{Mahadevan, Amala}} \yr{2016}  \at{The impact of submesoscale physics on primary productivity of plankton}.  \jt{Annual Review of Marine Science}  \bvol{8},  \pg{161--184}.

\bibitem[McWilliams(2016)]{2016_PRSA_McWilliams}
{\sc \au{McWilliams, James~C.}} \yr{2016}  \at{Submesoscale currents in the ocean}.  \jt{Proceedings of the Royal Society A: Mathematical, Physical and Engineering Science}  \bvol{472},  \pg{20160117}.

\bibitem[McWilliams(2021)]{Mcwilliams2021}
{\sc \au{McWilliams, James~C.}} \yr{2021}  \at{Oceanic frontogenesis}.  \jt{Annual Review of Marine Science}  \bvol{13},  \pg{227--253}.

\bibitem[Pham {\em et~al.\/}(2024)Pham, Verma, Sarkar, Shcherbina \& D’Asaro]{Pham2024}
{\sc \au{Pham, Hieu~T.}, \au{Verma, Vicky}, \au{Sarkar, Sutanu}, \au{Shcherbina, Andrey~Y.} \& \au{D’Asaro, Eric~A.}} \yr{2024}  \at{Rapid downwelling of tracer particles across the boundary layer and into the pycnocline at submesoscale ocean fronts}.  \jt{Geophysical Research Letters}  \bvol{51}.

\bibitem[Ponce {\em et~al.\/}(2019)Ponce, Van~Zon, Northrup, Gruner, Chen, Ertinaz, Fedoseev, Groer, Mao, Mundim {\em et~al.\/}]{ponce2019deploying}
{\sc \au{Ponce, Marcelo}, \au{Van~Zon, Ramses}, \au{Northrup, Scott}, \au{Gruner, Daniel}, \au{Chen, Joseph}, \au{Ertinaz, Fatih}, \au{Fedoseev, Alexey}, \au{Groer, Leslie}, \au{Mao, Fei}, \au{Mundim, Bruno~C} \& \au{others}} \yr{2019}  \at{Deploying a top-100 supercomputer for large parallel workloads: The niagara supercomputer}.  \bt{In {\em Proceedings of the Practice and Experience in Advanced Research Computing on Rise of the Machines (learning)\/}},  \pg{pp. 1--8}.

\bibitem[Ramadhan {\em et~al.\/}(2020)Ramadhan, Wagner, Hill, Campin, Churavy, Besard, Souza, Edelman, Ferrari \& Marshall]{Ramadhan2020}
{\sc \au{Ramadhan, Ali}, \au{Wagner, Gregory}, \au{Hill, Chris}, \au{Campin, Jean-Michel}, \au{Churavy, Valentin}, \au{Besard, Tim}, \au{Souza, Andre}, \au{Edelman, Alan}, \au{Ferrari, Raffaele} \& \au{Marshall, John}} \yr{2020}  \at{Oceananigans.jl: Fast and friendly geophysical fluid dynamics on gpus}.  \jt{Journal of Open Source Software}  \bvol{5},  \pg{2018}.

\bibitem[Samelson \& Skyllingstad(2016)]{samelson2016frontogenesis}
{\sc \au{Samelson, RM} \& \au{Skyllingstad, ED}} \yr{2016}  \at{Frontogenesis and turbulence: A numerical simulation}.  \jt{Journal of the Atmospheric Sciences}  \bvol{73}~(12),  \pg{5025--5040}.

\bibitem[Shakespeare(2016)]{Shakespeare2016}
{\sc \au{Shakespeare, Callum~J.}} \yr{2016}  \at{Nonhydrostatic wave generation at strained fronts}.  \jt{Journal of the Atmospheric Sciences}  \bvol{73},  \pg{2837--2850}.

\bibitem[Siegelman {\em et~al.\/}(2020)Siegelman, Klein, Thompson, Torres \& Menemenlis]{Siegelman2020}
{\sc \au{Siegelman, Lia}, \au{Klein, Patrice}, \au{Thompson, Andrew~F.}, \au{Torres, Hector~S.} \& \au{Menemenlis, Dimitris}} \yr{2020}  \at{Altimetry-based diagnosis of deep-reaching sub-mesoscale ocean fronts}.  \jt{Fluids}  \bvol{5}.

\bibitem[Smagorinsky(1963)]{smagorinsky1963general}
{\sc \au{Smagorinsky, Joseph}} \yr{1963}  \at{General circulation experiments with the primitive equations: I. the basic experiment}.  \jt{Monthly weather review}  \bvol{91}~(3),  \pg{99--164}.

\bibitem[Srinivasan {\em et~al.\/}(2023)Srinivasan, Barkan \& McWilliams]{Srinivasan2023}
{\sc \au{Srinivasan, Kaushik}, \au{Barkan, Roy} \& \au{McWilliams, James~C.}} \yr{2023}  \at{A forward energy flux at submesoscales driven by frontogenesis}.  \jt{Journal of Physical Oceanography}  \bvol{53},  \pg{287--305}.

\bibitem[Stone(1966)]{1966_JAS_Stone}
{\sc \au{Stone, Peter~H.}} \yr{1966}  \at{On non-geostrophic baroclinic stability}.  \jt{Journal of the Atmospheric Sciences}  \bvol{23},  \pg{390--400}.

\bibitem[Sullivan \& McWilliams(2018)]{2018_JFM_SullivanMcWilliams}
{\sc \au{Sullivan, Peter~P.} \& \au{McWilliams, James~C.}} \yr{2018}  \at{Frontogenesis and frontal arrest of a dense filament in the oceanic surface boundary layer}.  \jt{Journal of Fluid Mechanics}  \bvol{837},  \pg{341--380}.

\bibitem[Taylor \& Thompson(2022)]{2023_ARFM_TaylorThompson}
{\sc \au{Taylor, John~R} \& \au{Thompson, Andrew~F}} \yr{2022}  \at{Submesoscale dynamics in the upper ocean}.  \jt{Annual Review of Fluid Mechanics Annu. Rev. Fluid Mech. 2023}  \bvol{55},  \pg{103--127}.

\bibitem[Thomas(2012)]{Thomas2012}
{\sc \au{Thomas, Leif~N.}} \yr{2012}  \at{On the effects of frontogenetic strain on symmetric instability and inertia-gravity waves}.  \jt{Journal of Fluid Mechanics}  \bvol{711},  \pg{620--640}.

\bibitem[Thomas {\em et~al.\/}(2013)Thomas, Taylor, Ferrari \& Joyce]{Thomas2013}
{\sc \au{Thomas, Leif~N.}, \au{Taylor, John~R.}, \au{Ferrari, Raffaele} \& \au{Joyce, Terrence~M.}} \yr{2013}  \at{Symmetric instability in the gulf stream}.  \jt{Deep-Sea Research Part II: Topical Studies in Oceanography}  \bvol{91},  \pg{96--110}.

\bibitem[Trefethen {\em et~al.\/}(1993)Trefethen, Trefethen, Reddy \& Driscoll]{1993_Science_TrefethenTRD}
{\sc \au{Trefethen, Lloyd~N.}, \au{Trefethen, Anne~E.}, \au{Reddy, Satish~C.} \& \au{Driscoll, Tobin~A.}} \yr{1993}  \at{Hydrodynamic stability without eigenvalues}.  \jt{Science}  \bvol{261},  \pg{578--584}.

\bibitem[Verma {\em et~al.\/}(2019)Verma, Pham \& Sarkar]{2019_OM_VermaPS}
{\sc \au{Verma, Vicky}, \au{Pham, Hieu~T.} \& \au{Sarkar, Sutanu}} \yr{2019}  \at{The submesoscale, the finescale and their interaction at a mixed layer front}.  \jt{Ocean Modelling}  \bvol{140},  \pg{101400}.

\bibitem[Wienkers {\em et~al.\/}(2021{\natexlab{{\em a\/}}})Wienkers, Thomas \& Taylor]{Wienkers2021}
{\sc \au{Wienkers, A.~F.}, \au{Thomas, L.~N.} \& \au{Taylor, J.~R.}} \yr{2021{\natexlab{{\em a\/}}}}  \at{The influence of front strength on the development and equilibration of symmetric instability. part 1. growth and saturation}.  \jt{Journal of Fluid Mechanics}  \bvol{926}.

\bibitem[Wienkers {\em et~al.\/}(2021{\natexlab{{\em b\/}}})Wienkers, Thomas \& Taylor]{Wienkers2021_2}
{\sc \au{Wienkers, A.~F.}, \au{Thomas, L.~N.} \& \au{Taylor, J.~R.}} \yr{2021{\natexlab{{\em b\/}}}}  \at{The influence of front strength on the development and equilibration of symmetric instability. part 2. nonlinear evolution}.  \jt{Journal of Fluid Mechanics}  \bvol{926}.

\bibitem[Zemskova {\em et~al.\/}(2020)Zemskova, Passaggia \& White]{zemskova2020transient}
{\sc \au{Zemskova, Varvara~E}, \au{Passaggia, Pierre-Yves} \& \au{White, Brian~L}} \yr{2020}  \at{Transient energy growth in the ageostrophic eady model}.  \jt{Journal of Fluid Mechanics}  \bvol{885},  \pg{A29}.

\bibitem[Zhu {\em et~al.\/}(2024)Zhu, Yang, Li, Chen, Ma, Cai \& Wu]{Zhu2024}
{\sc \au{Zhu, Ruichen}, \au{Yang, Haiyuan}, \au{Li, Mingkui}, \au{Chen, Zhaohui}, \au{Ma, Xin}, \au{Cai, Jinzhuo} \& \au{Wu, Lixin}} \yr{2024}  \at{Observations reveal vertical transport induced by submesoscale front}.  \jt{Scientific Reports}  \bvol{14}.

\end{thebibliography}

\end{document}